\documentstyle[12pt]{article}

\def\cmm2{{\,\rm cm^{-2}}}
\def\cm2{{\,{\rm cm}^2}}
\def\cmm3{{\,{\rm cm}^{-3}}}
\def\gcmm3{{\,{\rm g\,cm^{-3}}}}

\def\fun#1#2{\lower3.6pt\vbox{\baselineskip0pt\lineskip.9pt
  \ialign{$\mathsurround=0pt#1\hfil##\hfil$\crcr#2\crcr\sim\crcr}}}

\begin{document}
\thispagestyle{empty}

\begin{center}
\rightline{FERMILAB--Pub--93/013-A}
\rightline{astro-ph/9305***}
\rightline{revised }

\vspace{0.6in}
{\Large\bf Testing for Gaussianity \\
 Through the Three
Point Temperature Correlation Function}\\

\vspace{.3in}
{\large\bf Xiaochun Luo and David N. Schramm} \\
\vspace{0.2in}

{ Departments of Physics and of Astronomy \& Astrophysics\\
Enrico Fermi Institute, The University of Chicago, Chicago, IL  60637-1433}\\

{ NASA/Fermilab Astrophysics Center,
Fermi National Accelerator Laboratory, Batavia, IL  60510-0500}\\

\end{center}

\vspace{.3in}

\centerline{\bf ABSTRACT}
\vspace{0.2in}

One of the crucial aspects of  density perturbations that are produced
by the standard inflation scenario is that they are Gaussian
where seeds produced by topological defects tend to be non-Gaussian.
The three point correlation
function of the temperature anisotropy of the cosmic microwave background
radiation (CBR) provides a sensitive test of this aspect of the primordial
density field. In this paper, this function is calculated in the general
context of  various allowed non-Gaussian models. It is shown that
by  COBE and the forthcoming  South Pole and Balloon CBR anisotropy data
may be able to test provide a crucial test of Gaussianity.
\vskip 0.3 in
\centerline{ PACS number: 98.80.Bp, 98.80.Dr}
\vfill\eject

Testing for the Gaussianity of the primordial fluctuation spectrum is of
critical importance to many cosmological models. In particular, traditional
cosmic
inflation \cite{inflation}
specifically predicts a Gaussian density fluctuation spectrum.
The scale invariant quantum fluctuations generated during the inflationary
epoch are expected
to  serve
as the primordial density perturbations which develop into the large scale
structures we observe today\cite{structure}.
 Competing models for structure
 formation, including  topological defects originating from cosmological
phase transitions \cite{defect}
and  non-standard inflation models\cite{nonGau}, will also
generate a scale invariant (or nearly scale invariant)
 power spectrum for density perturbations similar to that of inflation.
However, the statistics of these latter
 fluctuations are non-Gaussian. Thus, the Gaussianity of the fluctuations
provides a unique handle in discriminating different structure formation
scenarios. In this letter, we will discuss how to test this aspect of the
primordial density field
through the temperature anisotropy of the cosmic microwave background
radiation (CBR).

 As we showed \cite{luo1}, in
momentum space, the lowest order deviation from Gaussianity is described
by the bispectrum of the gravitational potential $\phi$, $P_{\phi} (k_{1},
k_{2}) = <\phi_{k_{1}}\phi_{k_{2}}
\phi_{-k_{1}-k_{2}}>$. When the perturbation is adiabatic so that the
temperature
anisotropy is related to the gravitational potential $\phi$ at the last
scattering surface through the Sachs-Wolfe \cite{sachs} formula:
\begin{equation}
{\delta T \over T} = {\phi \over 3},
\end{equation}
the three point temperature correlation function
is related to the bispectrum through
\begin{equation}
\xi_{T} (\hat{m}, \hat{n}, \hat{l}) = {1\over 27}\cdot\int
P_{\phi}(k_{1},k_{2},k_{3}){e^{i(\hat{k_{1}}\hat{m}+i\hat{k_{2}}\hat{n}
+i\hat{k_{3}}\hat{l})\eta_{0}}}\delta^{3}(\vec{k_{1}}+\vec{k_{2}}+\vec{k_{3}})
{d^{3}k_{1}d^{3}k_{2}d^{3}k_{3}\over(2\pi)^{9}},
\end{equation}
where $\eta_{0} = 2 H_{0}^{-1}$ is the distance to the last scattering surface,
and $\hat{m}, \hat{n}, \hat{l}$ are the beam directions.
  A non-vanishing three point function clearly indicates that the bispectrum is
not zero. Note that for Gaussian primordial perturbations, the bispectrum is
strictly zero in all cases. Thus, the
three point temperature correlation
function is a clean test of the Gaussianity of primordial fluctuations.

In \cite{toby}, Falk et al. found that in an inflationary model with cubic
self-interaction, $P_{\phi} (k_{1}, k_{2}, k_{3})$ is given by:
\begin{equation}
P_{\phi} = \beta (k_{1}k_{2} k_{3})^{-3} (k_{1}^{3} + k_{2}^{3} + k_{3}^{3}),
\end{equation}
where $\beta \sim 10^{-6}$.  In this paper,
 we will show that without invoking any new assumptions about inflationary
models, the non-linear gravitational evolution of the initial Guassian
perturbations will give rise to a three point correlation function, which
has a similar angular dependence to that which certain
 non-Gaussian inflationary models
predict, but a much larger amplitude than the one considered in \cite{toby}.
Then, we
extend the analysis in \cite{toby} to include more general cases of
 inflation, which produce not only the scale invariant but also the
``tilted'' perturbation spectrum. The extended analysis is helpful in
discussing
the effect of spectral index on the angular dependence of the three point
function.
To choose different non-Gaussian inflationary models through three point
temperature correlation function will be hard because of the gravitational
evolutionary effects. However,
the three point correlation function produced by a cosmological phase
transition
tends to have
 a distinctive angular dependence, which should enable one to
 prove or disprove the scenario through observations. Finally, we briefly
discuss the effect of noise
in the sky signal of CBR measurements on the analysis of three point
temperature correlation function.

  By taking into account  non-linear gravitational evolution, it is found that
there are two terms which contribute to the CBR temperature anisotropy:
\begin{equation}
{\delta T\over T} = {\phi \over 3} + 2 \int {\partial \phi\over \partial \eta
} d\eta,
\end{equation}
where the first term is the Sachs-Wolfe \cite{sachs} effect due to the
gravitational potential at the
last scattering surface, and the second term is the generalized
Rees-Sciama effect \cite{rees} due the
evolution  of the gravitational potential along the photon path.
When we adopt a flat cosmological model ($\Omega = 1$), the quasi-nonlinear
analysis gives \cite{roman,vishniac,luo1}:
\begin{equation}
\phi (k, \eta) = \phi_{i} (k) + a(\eta) \int J(\vec{k}, \vec{k^{'}},
\vec{k}-\vec{k^{'}}) \phi_{i}({k^{'}}) \phi_{i} (k-k^{'}),
\end{equation}
where $\phi_{i}$ is the gravitational potential at the last scattering surface,
$a(\eta) = ({\eta\over \eta_{i}})^{2}$ is the expansion factor of the universe
after decoupling and
\begin{equation}
J(\vec{k}, \vec{l}, \vec{m}) = 2 (\vec{l}\cdot \vec{m}) + {5(\vec{k}\cdot
\vec{l}) m^{2}\over k^{2}} + {5(\vec{k}\cdot\vec{m}) l^{2}\over k^{2}}.
\end{equation}
We first estimate the amplitude of the second term relative to the first
term: since the expansion factor $a$ after decoupling is $\sim (1 +
z_{dec})\sim 1000$, the
amplitude of the gravitational potential at the last scattering surface is
around $10^{-5}$ as suggest by COBE, thus  the ratio of the Rees-Sciama term
to the Sachs-Wolfe term is of order 0.01 - 0.1. As the non-linear effects
are contained in the Rees-Sciama term, it is this term that
contributes significantly to the
three point correlation function. For  comparison, the non-linear term
considered in \cite{toby} is $10^{-6}$ times smaller than the linear term.
This is a potential problem on testing inflationary models through
three point temperature correlation function. To be
observable the amplitude of the non-Gaussianity produced in these models has to
be large enough so that the gravitational
evolution  cannot completely dominate.

  The generic form of the bispectrum in inflationary models is given by:
\begin{equation}
P_{\phi}(k_{1}, k_{2}, k_{3})=\lambda [P_{\phi} (k_{1})
P_{\phi}(k_{2})+P_{\phi}(k_{2})P_{\phi}(k_{3})+P_{\phi}(k_{3})P_{\phi}(k_{1})],
\label{genric}
\end{equation}
where $\lambda$ is a constant and $P_{\phi} = <\phi(k) \phi(-k)>$ is related
to the power spectrum $P(k)$ simply through $P_{\phi}(k) = P(k) k^{-4}$.
The cubic self-interaction model corresponds to the case where
$\lambda \sim 10^{-6}$ with
  a scale invariant density perturbation spectrum, or
equivalently,
$P_{\phi} (k) \sim k^{-3}$; the non-linear gravitational
evolution effect corresponds to the case where $\lambda \sim 2(1 + z_{dec})/9
\sim 200$ and $P_{\phi} (k) \approx k^{-2}$.

Note that the two point temperature correlation function is
related to $P_{\phi}$ through:
\begin{equation}
C_{2} (\hat{m}, \hat{n}) = {1\over 9}\int P_{\phi}(k)e^{i\vec{k}(\hat{m} -
\hat{n})\eta_{0}}{d^{3}k\over(2\pi)^{3}}.
\end{equation}
The three point correlation function calculated from the bispectrum
given by Eq.\ref{genric}  is:
\begin{equation}
\xi_{T}(\hat{m}, \hat{n}, \hat{l}) = 3\lambda [C_{2}(\hat{m}, \hat{n})
C_{2}(\hat{n}, \hat{l}) + C_{2}(\hat{m}, \hat{n}) C_{2}(\hat{m}, \hat{l})
+ C_{2}(\hat{m}, \hat{l}) C_{2}(\hat{n},\hat{l})].
\end{equation}
This is a theoretical  relation between three point function and two point
function since the finite beam size effects haven't taken into account yet.
The formal treatment of the finite beam effect in CBR experiment can be found
 in \cite{silk,gouda}. The beam can be well approximated as
a Gaussian:
\begin{equation}
f(|\hat{m} -\hat{n}|, \sigma) = {1\over 2\pi\sigma^{2}} e^{-|\hat{m}
-\hat{n}|^{2}/2\sigma^{2}},
\end{equation}
and the observed temperature correlation function
will be the convolution of the theoretical correlation (infinite thin beam)
with the beam, which is
\begin {eqnarray}
 C_{3}(|\hat{m},\hat{n}, \hat{l}|, \sigma) =
\nonumber \\
\int  d\Omega_{1}^{'}
d\Omega_{2}^{'}d\Omega_{3}^{'} f(|\hat{m} -\hat{m^{'}}|, \sigma) f(|\hat{n}
-\hat{n^{'}}|
, \sigma)f(|\hat{l} -\hat{l^{'}}|, \sigma)
 C_{3}(|\hat{m^{'}},|\hat{n^{'}},
\hat{k^{'}}|, 0).
\end{eqnarray}
For a special configuration of three beams where $\hat{m}\cdot\hat{n} =
\hat{n}\cdot\hat{l} = \hat{l}\cdot\hat{m} = \cos\alpha$, the beam-smoothed
three point function is well approximated\cite{toby}
 as $[C_{2}(\cos\alpha|\sigma)]^{2}$
where $[C_{2}(\cos\alpha|\sigma)$ is the two point function with
 the monopole, dipole and quadruple terms removed.
Since the three point function is the products of two 2-point functions,
it has a stronger dependence on the power spectra index $n$. Mutipole expansion
of the 2-point function gives:
\begin{equation}
C_{2} (\hat{m}, \hat{n}) = \sum_{l} C_{l} (2l +1) P_{l} (\hat{m}\cdot\hat{n}).
\end{equation}
For a power law  spectrum $P(k) \sim k^{n}$, $C_{l}$ is given by
\begin{equation}
C_{l} ={1\over 5} ({Q_{rms}\over T_{0}})^{2}  {\Gamma({2 l+n-1}) \over \Gamma({
2l+5-n})} \cdot {\Gamma({9-n\over 2}) \over \Gamma({3-n\over 2})},
\end{equation}
where $Q_{rms}$ is the COBE measured quadruple\cite{cobe} and $
T_{0}$ is the black-body temperature of CBR. From the analysis of two
point correlation function, COBE can only put a loose bound on power
spectra index\cite{cobe}: $ n = 1.1 \pm 0.5$.  In Fig. 1, we plot
the three point function for two different power spectra: a scale
invariant $n=1$ spectrum and a ``tilted''
 spectrum where $n =0.7$. Notice how the three point
function depends strongly upon the power spectra index. Thus it is anticipated
 that
the analysis of the three point correlation function will put a more stringent
bound on $n$.

In order to test cosmological structure formation scenarios through the
three point temperature correlation function, we should have a clear
handle on what various models predict.  In the following,
we will show that the cosmological phase transition can
produce  distinctive angular dependences other than the form given above.

 Cosmological phase transitions are widely discussed in the context of
the structure formation \cite{kolb}. In the case of a
primordial phase transition,
the horizon size at the epoch of phase transition is small and topological
defects will form according to the Kibble mechanism \cite{kibble}. The
 analysis
of the three point correlation for defect-induced temperature anisotropy
depends crucially on the evolution of the defect-network and the work
along this line is still in progress. In this paper, we will show that
 it is instructive
to consider initially the three point function in the late-time phase
transition
 (LTPT) scenario \cite{hill}. The calculation is considerably simplified in
LTPT models  for the following reasons: (1) the last
 scattering surface is assumed
smooth in LTPT models, thus  temperature anisotropies
are solely   produced by the generalized Rees-Sciama effect,
\begin{equation}
{\delta T \over T} = 2 \int {\partial \phi \over \partial \eta} d\eta.
\end{equation}
since  the fluctuations in
density and gravitational potential are generated by the critical fluctuations
at the critical point of the phase transition, ${\partial \phi\over
\partial \eta} = \phi \delta (\eta - \eta_{p})$, where $\eta_{p}$ is the
conformal time at the phase transition point. Thus,  in
this latetime phase transition model, the temperature anisotropy takes
the following simple form:
\begin{equation}
{\delta T \over T}   = 2 \phi_{\eta = \eta_{p}}.
\end{equation}
(2)  For LTPT, the horizon
size is large so that the finite horizon-size effect is negligible.
 We can calculate the three point correlation function from  symmetry
considerations. As pointed out by Polyakov \cite{polyakov}, the
three point correlation function of the fluctuating field $\psi$ is completely
determined up to a dimensionless constant  by the conformal symmetry of the
system at the critical point. The explicit form for the three point function is
given by
\begin{equation}
\xi_{3} = < \psi(x_{1}) \psi(x_{2}) \psi(x_{3})>
= \eta c_{2}(x_{1}, x_{2}) c_{2}(x_{2}, x_{3}) c_{2}(x_{3}, x_{1}),
\end{equation}
where $c_{2} (x_{1}, x_{2}) = <\psi (x_{1})\psi (x_{2})>$ is the two
point function and $\eta$ is a constant.
In this letter, we assume that the gravitational potential $\phi$
is directly proportional to the underlying fluctuating field $\psi$. For this
case,  the three point temperature correlation function has the following
simple
relation to the two point function:
\begin{equation}
\xi_{T} (\hat{m}, \hat{n}, \hat{l}) = A \cdot C_{2} (\hat{m}, \hat{n})
C_{2} (\hat{n}, \hat{l}) C_{2} (\hat{m}, \hat{l}),
\end{equation}
where A is a dimensionless constant. The full beam-smearing effects
of the three point correlation function given above is messy and we will
report it elsewhere. However, in the special case when $\hat{m} \cdot \hat{n} =
\hat{n} \cdot \hat{l} = \hat{l}\cdot\hat{m} = \cos\alpha$,  it can be
approximated as $[C_{2} (\cos\alpha|\sigma)]^{3}$, where $C_{2}
(\cos\alpha|\sigma)$
is the 2-pt function with finite beam width $\sigma$, with monopole,
dipole and quadruple terms subtracted. We plot the approximated three point
function generated by the phase transition in Fig. (2).

  The result obtained from Eq. (9) \& Eq. (13) strongly suggest that
the general form of the three point function, expressed in terms of
 two point functions, is given by:
\begin{eqnarray}
\xi_{T}(\hat{m}, \hat{n}, \hat{l})=Q\cdot [C_{2}(\hat{m}, \hat{n})
C_{2}(\hat{n}, \hat{l}) + C_{2}(\hat{m}, \hat{n}) C_{2}(\hat{m}, \hat{l})
+ C_{2}(\hat{m}, \hat{l}) C_{2}(\hat{n},\hat{l})] \nonumber \\
+ A \cdot C_{2} (\hat{m}, \hat{n})
C_{2} (\hat{n}, \hat{l}) C_{2} (\hat{m}, \hat{l}),
\end{eqnarray}
where Q and  A are constants.
This is the archetype form of the three point correlation function
that the experimental analysis should be compared with.

The Gaussianity can be tested through the existing COBE and the forthcoming
South Pole and Balloon CBR anisotropy data by three point temperature
correlation function. In this letter, we have focused on the COBE data although
the idea and method discussed can equally apply to South Pole and Balloon
experiments.
The data set from  COBE DMR is especially suitable for carrying out this test.
On the one hand, the beam width of COBE is $7^{\circ}$, which is much larger
than the horizon size at decoupling ($\sim 2^{\circ}$). Most
non-linear causal processes which may lead to non-Gaussian signatures on
the cosmic microwave background (CMB)
 sky are smoothed out by the beam. On the other
hand, the COBE CMB map covers the whole sky. Thus,
  the boundary effects will be
minimized. However, the detected sky signal contain both the intrinsic
CBR temperature fluctuation and the instrumental noises,
\begin{equation}
{\delta T \over T}) _{obs} = { \delta T \over T})_{CBR} + {\delta T \over
T})_{noise}.
\end{equation}
The signal to noise ratio of the COBE data is 1:1 and
this is typical in all current CBR temperature anisotropy experiments. Thus,
it is important to consider the noise term seriously in the analysis
of the three point correlation function. Even if future analysis of the COBE
data do find a non-vanishing three-point temperature correlation, it may
due to the instrumental noise. However, if one adopts the usual assumption
about the noise term: (1) the noise is random Guassian noise which is not
correlated temporally or spatially; (2) the noise is not correlated with the
CBR signal, then
\begin{equation}
\xi_{obs} = \xi_{CBR},
\end{equation}
the three point correlation calculated from the raw observational data
will reflect directly the three point temperature correlation of the CBR,
 even if the noise term is comparable to the signal. This is
the another advantage
to using the  three point function to test Gaussianity of the initial
perturbations.

\bigskip
\centerline{Acknowledgements}
\bigskip

  We want to thank  Jim Fry, Michael Turner, Andrew Jaffe,
Albert Stebbins and especially Mark Srednicki for useful discussions. This work
is supported in part by NSF grant \#90-22629 and by NASA grant \#NAGW 1321 at
 the university of Chicago and by DOE and by NASA through grant \#NAGW 2381 at
Fermilab.

\vfill\eject
\centerline{FIGURE CAPTIONS}
\bigskip
\noindent
Fig. 1: The dependence of the 3-pt temperature correlation on the power
spectra index and angular separation. The solid line is for a scale
invariant density perturbation; the dash line is for a  ``tilted'' spectrum
with spectra index = 0.7.

\noindent
Fig. 2: The  angular dependence for the three point function
in different scenarios: the solid line represents the 3-pt function
generated by inflation;
the dash line represents the one generated by a cosmological phase
transition.
\vfill\eject
\end{document}